\documentclass[article]{JHEP3}

\usepackage{graphicx}
\usepackage{amsmath,epsfig}
\usepackage{amssymb,amsfonts}
\usepackage{latexsym}

\relax

\def\be{\begin{equation}}
\def\ee{\end{equation}}

\newcommand{\ha}{{1 \over 2}}

\newcommand{\bear}{\begin{eqnarray}}
\newcommand{\bea}{\begin{eqnarray}}
\newcommand{\eear}{\end{eqnarray}}
\newcommand{\eea}{\end{eqnarray}}
\def\hri#1#2{\href{http://arxiv.org/abs/#1}{[ArXiv:#1]#2}}
\def\hre#1#2{\href{http://arxiv.org/abs/#1/#2}{[ArXiv:#1/#2]}}

\newbox\pippobox

\def\II{\relax{\rm I\kern-.18em I}}

\def\l{\lambda}
\def\m{\mu}
\def\n{\nu}

\def\t{\theta}
\def\sp{\;\;\;,\;\;\;}

\def\tr{\ensuremath{\mathrm{Tr}}}
\def\str{\ensuremath{\mathrm{Str}}}
\def\t{\tau}
\def\al{A_{L}}
\def\ar{A_{R}}

\def\hal{\hat{A}_{L}}
\def\har{\hat{A}_{R}}

\def\ha{\hat{A}}
\def\hf{\hat{F}}

\def\L{\Lambda}
\def\cw{{\mathcal W}}
\def\ca{{\mathcal A}}
\def\cf{{\mathcal F}}

\title{Vector-axial vector correlators in weak electric field and the holographic dynamics of the chiral condensate}

\author{Ioannis Iatrakis$^a$, \href{http://hep.physics.uoc.gr/~kiritsis/}{Elias Kiritsis}$^{a,b}$\\
~\\
$^a$ \href{http://hep.physics.uoc.gr}{Crete Center for Theoretical Physics},
Department of Physics, University of Crete, 71003 Heraklion, Greece\\
~\\
$^b$ \href{http://www.apc.univ-paris7.fr}{APC, Universit\'e Paris 7}, \\ B\^atiment Condorcet, F-75205, Paris Cedex 13, France (UMR du CNRS 7164).}

\preprint{CCTP-2011-28}

\abstract{The transverse part of the vector-axial vector
  flavor current correlator in the presence of weak
  external electric field is studied  using holography. The correlator is calculated using a bottom-up model \href{http://arxiv.org/abs/arXiv:1003.2377}{proposed recently}, that includes both contributions of higher string states and the non-linear dynamics of the chiral condensate. It
  is shown that for low momenta the result agrees with the relation
  proposed by \href{http://arxiv.org/abs/arXiv:1010.0718}{Son and Yamamoto} motivated by a simpler holographic model. For large
  Euclidean momenta however, the two results are different.
  In the process, the difference of the vector and axial
  vector two point functions is also calculated. At large Euclidean momenta it is found
  that the first non-perturbative contribution, decreases as $q^{-6}$ as expected
  from QCD.
}

\begin{document}

\def\g{\gamma}
\def\go{\g_{00}}
\def\gi{\g_{ii}}

\maketitle %%%%%%%%%% THIS IS IGNORED %%%%%%%%%%%

\section{Introduction}
Global symmetries of classical field theories which fail to survive
quantization of the theory lead to quantum anomalies. These appear as
the non-invariance of the quantum effective action under the
symmetry transformation and as the violation of Ward identities
of certain correlators. One such example is the chiral (triangle) anomaly with one axial and two vector
currents. The longitudinal part of that correlator is not renormalized as was
shown by Adler and Bardeen, \cite{Adler:1969er}. However, the
transverse part is not necessarily constrained.

This correlator is of importance in
the context of the calculation of the two-loop
electroweak radiative corrections to the muon anomalous magnetic
moment, an important precision observable.
 In this context the transverse part of
the vector - axial vector QCD flavor current
correlator in the presence of weak electric field was studied in \cite{Knecht:2002hr} and \cite{Czarnecki:2002nt}.
 The physical significance of the above correlator led to its further
study both in perturbative QCD and in the strong coupling limit, in
\cite{Vainshtein:2002nv} and \cite{Knecht:2003xy}. A
non-renormalization theorem beyond the one-loop term was proved, perturbatively.

At the non-perturbative level, the correlator was studied using the operator product
expansion, see \cite{Knecht:2002hr}, \cite{Czarnecki:2002nt} and \cite{Vainshtein:2002nv}. There it was shown that the first most important non-perturbative correction
at high momenta scales like $1/q^6$ and is generated by the
expectation value of the dimension three (in the UV) antisymmetric
tensor current, $J^{\m\n}_{ij}\sim
\bar\psi^i_R\sigma^{\m\n}\psi^j_L$.
To be more precise, the general form of the
vector-axial vector correlator is, \cite{Vainshtein:2002nv}
\be
\begin{split}
\langle T \lbrace J_{\mu}^{a} (q) J_{\n}^{b \; (5)} (p) \rbrace
\rangle_{\hat F}&=-\tr({\mathcal Q}\, t^a t^b) {1 \over 4 \pi^2}(2 \pi)^4  \delta^{4}(q+p) \left\{ w_T(q^2)(-q^2
  {\mathcal O}_{\m\n} + q_{\m} q^{\sigma} {\mathcal O}_{\sigma \n}- q_{\n}
  q^{\sigma}{\mathcal O}_{\sigma \m}) \right.\\
&+\left. w_L (q^2) q_{\n}q^{\sigma}
  {\mathcal O}_{\sigma \m} \right\}
\label{2ptsepqcd1}
\end{split}
\ee
where ${\mathcal Q}$ is the electric charge matrix and $t^a$ are the flavor matrices.
The leading contribution comes from the dimension 2 operator
${\mathcal O}_{\m\n}= {1\over 2}\epsilon_{\m\n\rho\sigma} F^{\rho
  \sigma}$ and the coefficients read $w_L^{1 \; loop}=2 w_{T}^{1\;
  loop} =2 {N_c \over q^2}$, where $F^{\rho \sigma}$ is the electromagnetic field
strength tensor and $N_c$ the number of colors. In the chiral limit the next
contribution comes from non perturbative effects (non-trivial vevs) and in the large $N_c$
limit it reads
\be
{\mathcal O_{\mu\nu}}=\epsilon_{\mu \nu \rho \sigma} \langle \bar \psi_R \psi_L \rangle \langle \bar
\psi_R \sigma^{\rho \sigma} \psi_L \rangle
\ee
Its coefficient is computed at short distances and is $w_T^{non \;pert.}\sim{\alpha_s \over q^6}$, where $\alpha_s$ is the QCD fine structure constant. It should be emphasized that the
longitudinal part does not receive any correction beyond the one loop result.
Therefore, for the subleading corrections, one must have a non-perturbative setup in order to compute them.

In \cite{Son:2010vc} the simplest holographic bottom-up model for the meson sector was used to calculate the non-perturbative parts of the correlator,
(with partial success, as argued in \cite{Knecht:2011wh}). In
the simplest holographic model examined in \cite{Son:2010vc}, a relation of the transverse part of the
triangle anomaly to the vector and axial vector
two point functions was shown at all momenta,
Eq.(\ref{wtrellowq}) in section \ref{secvac}. A key point
of \cite{Knecht:2011wh} is that the holographic model considered in \cite{Son:2010vc}
does not contain contributions from higher-spin states, explaining the difficulty to find the
right subleading terms in the operator product expansion at large
Euclidean momentum. However, it should be noted that despite the mismatch at high momenta, the proposal is very important at low momenta and seems to pass several tests with data.

In the present work, we revisit the axial correlator in question, and its calculation from holography. We will use a setup that contains contributions from higher spin states as well model the dynamics of the chiral condensate in a more realistic fashion. We will find that the low energy structure of the transverse correlator is robust and agrees with Son-Yamamoto, but the high-energy structure is affected by all the extra ingredients that also appear in QCD. Although the $1/q^6$ is absent in our model, this is due to the unrealistic theory of glue we have used for simplicity.
On the other hand, the model has the correct subleading behavior at large $q$ for the difference $\Pi_V(q)-\Pi_A(q)$.

To motivate the setup it is important to revisit the low-dimension operators (dimension=3) in the flavor sector and their realization in string theory.
At the spin-zero level we have the (complex) mass operator
\be
\bar\psi^i_R\psi^j_L\leftrightarrow  T_{ij}
\ee
dual to a complex scalar transforming as $(N_f,\bar N_f)$ under the flavor symmetry $U(N_f)_R\times U(N_f)_L$.
At the spin-one level we have the two classically conserved currents
\be
\bar\psi^i_L\sigma^{\m}\psi^j_L\leftrightarrow A^{\m}_{L,ij}\sp \bar\psi^i_R\bar\sigma^{\m}\psi^j_R\leftrightarrow
 A^{\m}_{R,ij}
\ee
They transform in the adjoint of the $U(N_f)_R$ respectively the $U(N_f)_L$ symmetry.
The flavor symmetry is expected to arise in string theory from $N_f$ flavor branes (R) and $N_f$ favor antibranes (L).
The precise realization and dimensionality of the branes depends on the theory. In the most popular
top-down theory of Sakai and Sugimoto, \cite{ss} the flavor branes are 8-dimensional (while the full bulk is 10-dimensional and the gauge theory 5 dimensional) while in a 5-dimensional setup
expected to hold for the minimal YM realization, the favor branes are expected to be space-filling $D_4$ branes, \cite{ihqcd}.
Due to the quantum numbers, the vectors are the lowest modes of the fluctuations of the open strings with both ends on the D branes
($A^{\m}_{R}$), or the anti-D branes , $A^{\m}_{L}$.

The bifundamental  scalar T, on the other hand,  is the lowest mode of the $D-\bar D$ strings, compatible with its quantum numbers.
Its holographic dynamics is dual to the dynamics of the chiral condensate.
This is precisely the scalar that in a brane-antibrane system in flat space is the tachyon whose dynamics has been studied
profusely in string theory, \cite{sen}.  It has been proposed that the non-linear DBI-like actions proposed by Sen and others
are the proper setup in order to study the holographic dynamics of chiral symmetry breaking, \cite{ckp}.
This dynamics was analyzed in a toy example, \cite{ikp1,ikp2}, improving several aspects of the hard \cite{hard}, and soft wall models, \cite{soft}. We will keep referring to $T$ as the ``tachyon", as it indeed corresponds to a relevant operator in the UV.

Going further,  the antisymmetric current
\be
\bar\psi^i_R\sigma^{\m\n}\psi^j_L\leftrightarrow  B_{\m\n}^{ij}
\ee
is dual to a two-index antisymmetric
tensor\footnote{Recent discussions on the inclusion of this antisymmetric tensor explicitly in the holographic flavor action can
be found in \cite{anti}.}    that transforms as $(N_f,\bar N_f)$ under the flavor group.
It therefore originates in the $D-\bar D$ sector, and is a stringy descendant of the tachyon.
Indeed, in flat-space open-string spectra, the antisymmetric tensor appears at the level just above the tachyon, arising from two antisymmetrized oscillators acting
on the ground state, with reversed  GSO projection\footnote{Such stringy states were recently discussed in connection with the Sakai-Sugimoto model and mesons in  \cite{ss2}.}.

One of the characteristics of the Sen action for the tachyon, is its non-linearity, and the fact that contains (a class of) higher derivatives. In analogy with the standard DBI action, it contains contributions from all intermediate open string states, to leading order in derivatives.
In the open string sector, leading order implies that the effective action for the tachyon scalar is a function of its first derivatives but not higher ones.
Therefore, we expect that the DBI-like Sen action contains corrections due to the stringy modes and in particular the antisymmetric tensor mode.

Therefore this, and the fact that chiral symmetry breaking   is dynamical makes an evaluation of the axial correlator in such holographic models interesting.

In this paper we will
calculate this correlator  using the
simplified holographic model proposed and analyzed in \cite{ikp1},
\cite{ikp2}. The relation (\ref{wtrellowq}) derived  in \cite{Son:2010vc} in the simpler model is
not valid in general here.  However we find that it is approximately  valid  for sufficiently low momenta in our model. When on the other hand the momentum is well above the QCD scale then
the two quantities differ.

We have examined the large $q^2$ behavior of $w_T$. We have found numerically that  the first non-perturbative correction falls off as $1/q^8$. We conclude that the dimensions six operators somehow cancel in this background.
We have also calculated the large momentum behavior of $\Pi_{V}-\Pi_A$ and find that it asymptotes to $1/q^6$ as expected in QCD. This corrects our analytic estimate in \cite{ikp2}.

We have also calculated $w_T$ as a function of the bare quark mass. Using our earlier fits to the meson spectrum \cite{ikp2}, we calculate the correlator for two non-zero masses. The first is the up-down quark mass, where we observe that the correlator is almost identical to that with $m_q=0$.
We also calculate it with a mass matching the strange quark mass, and find that it is larger as shown in figure \ref{wsmallqcorr2}.

We have good reasons to expect that the same open string sector with a proper glue sector, close to QCD, as for example in \cite{ihqcd}, will provide a reliable correlator both at high and low momenta.

\section{Action and dynamics of holographic chiral symmetry breaking}

We consider a system of $N_f$ pairs of $D_4$-branes - $\bar
D_4$-antibranes in a fixed bulk gravitational background that describes the ``glue". The background was analyzed in
\cite{ks1} and \cite{ks2} and is a solution of the non-critical string theory action in six dimensions. The metric is an $AdS_6$
soliton

\be
ds^2 \equiv-g_{tt}dt^2+g_{zz}dz^2+g_{xx}dx_{3}^2+g_{\eta\eta}d\eta^2=\frac{R^2}{z^2} \left[ dx_{1,3}^2 +  f_\Lambda^{-1} dz^2 + f_\Lambda\, d\eta^2 \right]
\ee
where $f_{\Lambda}=1-{z^5 \over z_{\Lambda}^5}$  and the $\eta$ direction is
cigar shaped with its tip to be at $z_{\Lambda}$, so $z \in
[0,z_{\Lambda}]$. There is also a constant dilaton and a RR-form which is

\be
F_{(6)}=Q \sqrt{-g_{(6)}} d^6 x
\label{bbrrform}
\ee
$Q$ is a constant which will be fixed by matching to the anomaly of
the dual boundary field theory as it was proposed in
\cite{Witten:1998qj} and further analysed in \cite{Freedman:1998tz}.
Although the bulk geometry is not very close to standard YM, compared to finer constructions like \cite{ihqcd}, it has the advantage of simplicity, and this is the main reason that we consider it here. It is the bulk geometry obtained from a five-dimensional supersymmetric CFT compactified on a ``small" circle, with supersymmetry breaking boundary conditions for the fermion operators.   Despite the simplicity of the geometry, it turns out to fulfill the main qualitative necessary ingredients.

We now consider the generalization of Sen's action \cite{Sen:2003tm} for
describing $N_f$ coincident pairs of $D_4$-branes - $\bar
D_4$-antibranes, see \cite{ckp} and references therein. For the present study, the full non-abelian "tachyon-DBI"
is not necessary. The non abelian results that we need, come from  a
simple generalization of the abelian ones, which were found in
\cite{ckp} and \cite{ikp2}. The $D_4$ - $\bar D_4$ pairs are taken at a fixed point in $\eta$ direction\footnote{This is because in the 5-dimensional world, this is the case that is expected to describe the flavor sector.}. The action is

\be
S= - \int d^4x dz\, \str \left[ V(|T|)
\left(\sqrt{-\det {\bf A}_L}+\sqrt{-\det {\bf A}_R}\right) \right]
\label{generalact}
\ee
where
\be
{\bf A}_{(i)MN}=g_{MN} + {2\pi \alpha' \over g_{V}} {\mathcal F}^{(i)}_{MN}
+ \pi \alpha' \lambda \left((D_M T)^* (D_N T)+
(D_N T)^* (D_M T)\right) \;\;, M,N=1,\ldots,5
\label{Senaction}
\ee
$(i)=L,R$ denotes the left and right parts of the $U(N_f)_{L/R}$ gauge field
strengths ${\mathcal F}=d {\mathcal A}- i \ca \wedge \ca$.
More details about the definitions and conventions can be found in \cite{ckp}.

The tachyon
$T$ is a complex bifundamental scalar. We also define
\be
D T \equiv d T + i T {\mathcal A_L}- i {\mathcal A_R} T
\ee
We introduced the couplings $g_{V}$ and $\lambda$ which determine the
normalization of the bulk fields. These two couplings can be fixed by
matching the results of the vector and scalar two point functions as
calculated in the bulk on the one hand and in QCD on the other
hand, as was proposed in \cite{Erlich:2005qh} and \cite{Da Rold:2005zs}.
They have been matched in \cite{ikp2}.

The tachyon potential is
\be
V(\tau)= {\cal K}\, e^{-\frac12 \m^2 \tau^2}
\label{tachyonpot}
\ee
where ${\cal K}$ is a constant. The vacuum of the
above theory was analyzed in \cite{ikp2}. The only nonzero
field is the tachyon which diverges at the tip of the cigar. As it is pointed
out in \cite{Kruczenski:2003uq} and \cite{ckp}, in case that the $N_f$
quarks have the same mass the vacuum of the non abelian action consists of
$N_f$ copies of the abelian solution

\be
\langle T \rangle = \tau(z) \mathbb{I}
\label{tau1}
\ee
The background solution for $\tau(z)$ was studied in \cite{ikp2}. The
near-boundary expansion ($z\to 0$) of the tachyon reads

\be
\t(z)=c_1 z + {\m^2 \over 6} c_1^3 z^3 \log z +c_3 z^3 + \ldots
\ee
where $c_1$ is proportional to the quark mass and $c_3$ is proportional
to the vacuum expectation value of the $\bar q q$ operator.

The study of the above model in \cite{ikp1}, \cite{ikp2} led to
the description of many low energy QCD properties.

\begin{itemize}

\item The model incorporates  confinement in the sense that the quark-antiquark potential
computed with the usual AdS/CFT prescription \cite{Sonnenschein:1999if} confines.
Moreover, magnetic quarks are screened.
The background solution stems from a gravitational action, that allows, for instance,
to compute thermodynamical quantities. All of this are properties associated to the
background geometry and were already discussed in  \cite{ks2}.

\item The string theory nature of the bulk fields dual to the quark bilinear currents is readily identified:
they are low-lying modes living in a brane-antibrane pair.

\item Chiral symmetry breaking is realized dynamically and consistently, because of the
tachyon dynamics.
See \cite{Sui:2009xe} for discussion and possible solutions in the soft-wall model
context.

\item In this model, the mass of the $\rho$-meson grows with increasing quark mass, or,
more physically, with increasing pion mass. This welcome physical feature is absent in the soft wall model, \cite{soft}.
It occurs here  because the tachyon potential multiplies the full action and in particular
the kinetic terms for the gauge fields, which therefore couple to the chiral symmetry breaking vev.
In our previous work \cite{ikp1}, we exploited this fact in order to fit the strange-strange mesons
together with the light-light mesons, with rather successful results.
In \cite{Shock:2006qy}, the authors added the strange quark mass to the hard
wall model and computed the dependence of vector masses on the quark mass. In that case however,  this dependence
of the vector masses originated  only from the non-abelian structure and therefore misses at
least part of the physics\footnote{On the other hand, the quark mass dependence of the $\rho$-meson
can be seen in different top-down models, see \cite{cdk} for a recent work in the context of the
Klebanov-Strassler model.}.

\item The soft wall requires assuming a quadratic dilaton in the closed string theory background.
It has been shown that such a quadratic dilaton behaviour can never be derived from a gravitational action while keeping
the geometry to be that of AdS.\footnote{This was shown in  \cite{ihqcd2}. In
\cite{ihqcd} such behavior can be implemented for glue, but the metric changes appropriately, an important ingredient for implementing confinement in the glue sector.}.
That the background is not found as a solution is a shortcoming if for instance one wants to study the
thermodynamics of the underlying glue theory. The thermodynamics of the soft wall model is therefore ill-defined.
 In the present  model, the background is a solution
of a two-derivative approximation to non-critical string theory.
In order to obtain Regge behaviour, we also needed a further  assumption: that the
tachyon potential is asymptotically gaussian. However, this is rather natural since this potential has appeared
in the literature, for instance \cite{Kutasov:2000aq,Takayanagi:2000rz}.

\item Considering that the dynamics is controlled by a tachyon world-volume action automatically
provides the model with a WZ term of the form given in
\cite{Kennedy:1999nn,Kraus:2000nj,Takayanagi:2000rz,oz}. In \cite{ckp} it was shown that properties like discrete symmetries (parity and charge
conjugation)
and anomalies are,
in general,
correctly described by analyzing this term and we will detail them also here.

\end{itemize}

In practice,  the
matching of the the predicted mass spectrum and some decay
constants to experimental data is considered successful since
the model has two parameters which correspond to those of QCD  (quark mass and
QCD scale) and one phenomenological parameter which is fitted to
data\footnote{It turns our that spectra and decay constants depends rather weakly on this phenomenological parameter, \cite{ikp2}.}. Eventually, the rms error of numerical fits to spectra and decay constants is $10\% - 15\%$. In conclusion this simple bottom-up model,
which is string theory inspired, incorporates many interesting QCD features.

\section{The Wess Zumino action}
The Wess-Zumino term describing the coupling of the flavor
branes-antibranes to the RR background field was studied in
\cite{Takayanagi:2000rz}, \cite{Kennedy:1999nn}, \cite{Kraus:2000nj}
and is

\be
S_{WZ}=T_{4} \int_{\mathcal M_{p+1}} C\wedge \str\, e^{i 2 \pi
  \alpha' \mathcal{\bf F}}
\label{wzact}
\ee
where $C=\sum_n (-i)^{p-n+1 \over 2} C_n$ is a sum of the RR fields, and
$\mathcal{\bf F}= d \mathcal{\bf A} -i \mathcal{\bf A} \wedge
\mathcal{\bf A}$. The integration picks up the $(p+1)$- form of the
infinite sum of forms. In terms of the tachyon field and the left and right $U(N_f)$ gauge fields we have

\be
i\mathcal{\bf A}=\left(\begin{array}{cc} i\ca_L & T^\dagger\\
T & i\ca_R \end{array}\right)\,,\qquad
i\mathcal{\bf F}=\left(\begin{array}{cc} i\cf_L-T^\dagger T & DT^\dagger\\
DT & i\cf_R-TT^\dagger\end{array}\right)
\label{AFdef}
\ee
We also set $2 \pi \alpha'=1$.
The Wess-Zumino action on the worldvolume of the $D_4-\bar D_4$ flavor
branes in the background of $N_c$ $D_4$ color branes is

\be
S_{WZ}=i T_4 \int_{\mathcal M_{5}} C_{-1} \wedge \left. \mathrm{Str} \, e^{i {\mathcal F}}\right|_{6\mathrm{-form}}=
i T_4 \int_{\mathcal M_{5}} F_{(0)} \wedge \Omega_5
\label{wz04}
\ee
where $F_0= d C_{-1}$ is proportional to the number of colors. We work
in a non-critical supergravity 6-dimensional background which has a
non trivial RR form with field strength given in (\ref{bbrrform}).
It's dual form is $F_{(0)}=\star F_{(6)}=Q$, so

\be
S_{WZ}=i T_4 Q \int \Omega_5
\ee
where $i {T_4 Q=i {N_c \over 4 \pi^2}}$ was fixed by matching to the
QCD chiral anomaly in \cite{ckp}.  $\Omega_5$ is the 5-form that comes
from the expansion of $e^{i {\cal F}}$ and was found in \cite{ckp}

\be
\begin{split}
\Omega_5 &= {1\over 6} \tr\, e^{-\t^2} \left\{-i \ca_L \wedge \cf_L \wedge
    \cf_L+{1\over 2} \ca_L\wedge \ca_L \wedge \ca_L \wedge \cf_L +{ i \over 10}
    \ca_L \wedge \ca_L \wedge \ca_L \wedge \ca_L \wedge \ca_L \right. \\
&+ i \ca_R \wedge \cf_R
  \wedge \cf_R -{1\over 2} \ca_R \wedge \ca_R \wedge \ca_R \wedge \cf_R -{i
    \over 10} \ca_R \wedge \ca_R \wedge \ca_R \wedge \ca_R \wedge \ca_R \\
&+\t^2 \Big[ i \ca_L \wedge \cf_R \wedge \cf_R -i \ca_R \wedge \cf_L \wedge \cf_L
  +{i \over 2} (\ca_L - \ca_R) \wedge (\cf_L \wedge \cf_R +\cf_R \wedge \cf_L)
 \\
&+{1 \over 2} \ca_L \wedge \ca_L \wedge \ca_L \wedge \cf_L -{1\over 2} \ca_R
  \wedge \ca_R \wedge \ca_R \wedge \cf_R+ {i \over 10} \ca_L \wedge \ca_L \wedge
\ca_L \wedge \ca_L \wedge \ca_L \\
&- {i \over 10} \ca_R \wedge \ca_R \wedge \ca_R \wedge \ca_R \wedge \ca_R \Big]\\
&+ i \t^3 d\t \wedge \Big[ (\ca_L \wedge \ca_R -\ca_R \wedge \ca_L)\wedge
  (\cf_L+\cf_R) + i \ca_L \wedge \ca_L \wedge \ca_L \wedge \ca_R \\
&-{i \over 2} \ca_L \wedge \ca_R \wedge \ca_L \wedge \ca_R + i \ca_R \wedge \ca_R
\wedge \ca_R \wedge \ca_R \Big] \\
&\left. +{i \over 4} \t^4 (\ca_L-\ca_R) \wedge (\ca_L-\ca_R) \wedge (\ca_L-\ca_R) \wedge (\ca_L-\ca_R) \wedge (\ca_L-\ca_R)
\right\}
\end{split}
\label{o5full}
\ee
We now split the $U(N_f)$ gauge fields into their $U(1)$ and $SU(N_f)$
parts

\be
\ca_L={\hal \over \sqrt{2 N_f}}+\al \sp \ca_R = {\har \over \sqrt{2 N_f}}+ \ar
\ee
where we denote as $\al, \ar$ the $SU(N_f)$ part and $\hal, \har$ the
$U(1)$ part of the gauge field. We also consider the vector combination of the $U(1)$
fields, $\hal=\har=\ha$.
Since, we are interested in the calculation of the correlation
function of the vector and the axial current in the presence of a weak
electromagnetic field we expand $\Omega_5$ to linear order in $\hf$
and quadratic in the $SU(N_f)$ gauge fields. So, the relevant terms of
$\Omega_5$ are

\be
\begin{split}
\Omega'_5 &= {1\over 6} \tr\, e^{-\t^2} \left\{-i \ca_L \wedge \cf_L \wedge
    \cf_L +i \ca_R \wedge \cf_R \wedge \cf_R +\t^2 \Big[ i \ca_L
    \wedge \cf_R \wedge \cf_R - i \ca_R \wedge \cf_L \wedge \cf_L \right. \\
&\left.  +{i \over 2} (\ca_L - \ca_R) \wedge (\cf_L \wedge \cf_R +\cf_R
  \wedge \cf_L) \Big] + i \t^3 d\t \wedge (\ca_L \wedge \ca_R -\ca_R \wedge \ca_L)\wedge
  (\cf_L+\cf_R) \right\}\\
\end{split}
\ee
 We also define the vector and axial vector fields
\be
 V^a={\al^a +\ar^a \over 2} \sp A^a={\al^a - \ar^a \over 2}
\ee
where $a=1, \ldots, N_f^2-1$.
After the decomposition the action reads
\be
\begin{split}
S^{lin.}_{WZ}&={N_{c} \over 8 \pi^2} \int_{\mathcal M_5}
e^{-{1\over 2}\m^2  \tau^2}\hf \wedge \tr\, (\al-\ar) \wedge d (\ar+\al)\\
&={N_{c} \over 2\pi^2} \int_{\mathcal M_5}
e^{-{1\over 2}\m^2  \tau^2}\hf \wedge \tr \,(A \wedge d V)\\
\label{finalwz}
\end{split}
\ee
In order to express the action in the above form, we have added some boundary terms
to the initial action

\be
S_{WZ}^{lin.}= i {N_c \over 4 \pi^2} \int \Omega'_5+ S_{b. \, 1} +
S_{b. \, 2}+S_{b \, 3}
\ee
where

\be
\begin{split}
S_{b.\, 1} & ={N_{c} \over 24 \pi^2} \int_{\partial \mathcal
  M_5} e^{-{1 \over 2}\m^2  \t^2} \ha \wedge \tr\, (\ar \wedge d\ar
-\al \wedge d \al -3 \al \wedge d \ar +3 \ar \wedge d \al) \\
S_{b.\, 2} & ={N_{c} \over 24 \pi^2} \int_{\partial \mathcal M_5}
{\m^2\over 2} \t^2 e^{-{1 \over 2}\m^2  \t^2} \ha \wedge \tr\, (\ar \wedge d\ar
-\al \wedge  d \al -3 \al \wedge d \ar +3 \ar \wedge d \al) \\
S_{b.\, 3} & =-{N_{c} \over 24 \pi^2} \int_{\partial \mathcal M_5}
\m^2 \t^2 d \left( e^{-{1 \over 2}\m^2 \t^2} \right) \wedge \ha \wedge \tr\, ( \al \wedge \ar )\\
\end{split}
\label{bount}
\ee

\subsection{The chiral anomaly in the presense of the condensate}

We will now calculate the anomaly under a  $U(N_f)_V$ symmetry transformation in the presence
of a non zero tachyon (=chiral vev). In the derivation of the
expression (\ref{o5full}), the tachyon was considered to be proportional to unity matrix (\ref{tau1}), which
breaks the flavor symmetry to its diagonal subgroup, $U(N_f)_L\times
U(N_f)_R \rightarrow U(N_f)_V$. Hence we can only test transformations that preserve this form
,namely vector transformations.
The variation of the action (anomaly) under the symmetry transformation can be  written as
\be
(D_{\mu} \langle J^{\m}(x)\rangle)_a=\left. \left( D_{\mu} {\delta \cw
  \over \delta A_{\m}(x)}\right)_a\right|_{A=0}=-{\mathcal A_a (x)}
\label{andef}
\ee
where $J^{\m}$ is the symmetry current and $\cw$ is the generating
functional of the theory. Then, the variation under the symmetry gives

\be
\delta_{\Lambda} \cw[A]=-\delta_{\Lambda} S_{bulk} [A]=\int d^{4} x \L^a {\mathcal A}_a
\label{anact}
\ee
where $\Lambda$ is the transformation parameter. Direct calculation gives,

\begin{equation}
\begin{split}
\delta S_{WZ}=- {i N_c \over 24 \pi^2} \int_{\partial {\mathcal M_5}\,
  \epsilon}
&\tr\;
\Lambda \Big\{\,e^{-{\mu^2 \over 2}\tau(\epsilon)^2} (1+{\mu^2 \over 2}\tau(\epsilon)^2) \big[-\cf_L\wedge \cf_L-
\frac{i}{2}(\ca_L\wedge \ca_L\wedge \cf_L+\\
&+\ca_L\wedge \cf_L\wedge \ca_L+\cf_L\wedge \ca_L\wedge \ca_L)+\frac{1}{2} \ca_L\wedge \ca_L\wedge
\ca_L\wedge \ca_L+\cf_R\wedge \cf_R +\\
&\frac{i}{2}(\ca_R\wedge \ca_R\wedge \cf_R+\ca_R\wedge \cf_R\wedge \ca_R+\cf_R\wedge \ca_R\wedge \ca_R)-\frac{1}{2}\ca_R\wedge \ca_R\wedge \ca_R\wedge \ca_R\big]-\\
&-{\mu^2\over 4} \tau(\epsilon)^2\,d e^{-{\mu^2 \over 2} \tau(\epsilon)^2} \wedge \big[(\ca_L-\ca_R)\wedge (\cf_L+\cf_R)+(\cf_L+\cf_R)
\wedge (\ca_L-\ca_R)+\\
&i(\ca_L\wedge \ca_L\wedge \ca_L-\ca_L\wedge \ca_L\wedge \ca_R-\ca_R\wedge \ca_L\wedge \ca_L+\ca_L\wedge \ca_R
\wedge \ca_R+\\
&+\ca_R\wedge \ca_R\wedge \ca_L-\ca_R\wedge \ca_R\wedge \ca_R)\big]\Big\}
\label{anomaly}
\end{split}
\end{equation}
where $\epsilon$ is the UV cut off near the AdS boundary. We have also used the following variations of the fields
\be
\begin{split}
&\delta_\Lambda \ca_{L/R}=-i\, D\Lambda=-i\,d\Lambda -\ca_{L/R} \Lambda+\Lambda \ca_{L/R}\\
&\delta_\Lambda \cf_{L/R} =[\Lambda,\cf_{L/R}]
\end{split}
\ee

In case that the tachyon is a function of the radial
AdS coordinate only, so the quark condensate and mass are not spacetime
 dependent, the term which is proportional to
$d e^{-{\mu^2 \over 2} \tau(\epsilon)^2}$ in (\ref{anomaly}) is
zero. Hence, we recover the known QCD flavor anomaly up to an overall
consatnt term which depends on the UV cutoff, and can be reabsorbed in the
coupling, $T_4$ of the Wess Zumino action, Eq.(\ref{wzact}).
We notice that the anomaly depends on the condensate in case that
there is a finite UV cutoff and the condensate has non trivial
spacetime dependence. But also in this case, when we remove the cut-off (take $\epsilon
\rightarrow 0$), Eq.(\ref{anomaly}) reproduces the known QCD anomaly, see \cite{ckp},

\begin{equation}
\begin{split}
\delta S_{WZ}=- {i N_c \over 24 \pi^2} \int_{\partial {\mathcal M_5}}
&\tr\; \Lambda \Big\{-\cf_L \wedge \cf_L-\frac{i}{2}(\ca_L\wedge
\ca_L\wedge \cf_L +\ca_L\wedge \cf_L\wedge \ca_L+\cf_L\wedge
\ca_L\wedge \ca_L) + \\
&\frac{1}{2} \ca_L\wedge \ca_L\wedge
\ca_L\wedge \ca_L+\cf_R\wedge \cf_R +\frac{i}{2}(\ca_R\wedge
\ca_R\wedge \cf_R+ \\
&\ca_R\wedge \cf_R\wedge
\ca_R+\cf_R\wedge \ca_R\wedge \ca_R)-\frac{1}{2}\ca_R\wedge \ca_R\wedge \ca_R\wedge \ca_R \Big\}
\end{split}
\label{anomalyfinal}
\end{equation}

 In the case of the linearized action in the presence of the external EM field, (\ref{finalwz}), that we use
here, Eq.(\ref{finalwz}), it is observed that there is no triangle anomaly
under $U(1)_V$ and $SU(N_f)_V$, due to the addition of the boundary
terms which are given in (\ref{bount}). This is in agreement with
treatment of anomalies in quantum field theory, where one may consider
the one loop (due to fermion loops) effective action of the theory, which is a
functional of external vector and axial vector fields, and add local
counterterms in order to cancel the vector anomalies and be left
with the axial vector anomaly. This form of the anomaly is called the
Bardeen anomaly.

\section{The Vector-Axial vector correlator}
\label{secvac}

To calculate the vector-axial vector correlator in the
presence of an external weak electromagnetic electric field
we use the gauge, where $V_z^a=A_z^a=0$ and the split of
the relevant fields in momentum space is

\be
\begin{split}
V^a_{\mu}(q,z)&=P^{\bot \n}_{\m}(q) V_{\n}^{a \, (0)}(q) \psi^V(q,z)\\
A^a_{\mu}(q,z)&=P^{\bot \n}_{\m}(q) A_{\n}^{a \, (0)}(q) \psi^A(q,z)-P^{\| \,
\n}_{\mu}(q)  A_{\n \, (0)}^{a}(q) \phi(z)\\
\label{fieldsplit}
\end{split}
\ee
where the projection operators read $P^{\bot \n}_{\m}(q)=(\delta_{\m}^{\n} -{q_{\m}q^{\n} \over
  q^2})$, $P^{\| \,\n}_{\mu}(q)={q_{\m} q^{\n} \over q^2}$.
The correlator is defined as

\be
\langle T \lbrace J^{\mu}_{a} (x) J^{\n \; (5)}_{b} (y) \rbrace
\rangle_{\hat F}
= {\delta^2 S^{lin.}_{WZ}[A] \over \delta V_{\m}^{a \, (0)} (x) \delta
  A_{\n}^{b \, (0)} (y)}
\ee
In general, the correlator can be split to a transverse and a
longitudinal part. So, to linear order in $\hf$ it reads

\be
\begin{split}
\langle T \lbrace J_{\mu}^{a} (q) J_{\n}^{b \; (5)} (p) \rbrace
\rangle_{\hat F}&=-\tr({\mathcal Q}\,  t^a t^b) {1 \over 4 \pi^2}(2 \pi)^4  \delta^{4}(q+p) \left\{ w_T(q^2)(-q^2 \tilde
  F_{\m\n} + q_{\m} q^{\sigma} \tilde F_{\sigma \n}- q_{\n}
  q^{\sigma}\tilde F_{\sigma \m}) \right.\\
&+\left. w_L (q^2) q_{\n}q^{\sigma} \tilde
  F_{\sigma \m} \right\}
\label{2ptsep}
\end{split}
\ee
where $\tilde F_{\m \n}={1 \over 2} \epsilon_{\m \n \rho \sigma}
\hf^{\rho \sigma}$. Substituting  (\ref{fieldsplit}) in the
action (\ref{finalwz})

\be
\begin{split}
S^{lin.}_{WZ}&= -{\tr({\mathcal Q}\,  t^a t^b) \over 2}{N_c \over 4 \pi^2} \int {d^4 q \over
  (2\pi)^4} \int dz \,e^{-{1\over 2} \m^2 \t(z)^2} \epsilon^{\mu \nu z\rho \sigma}
\hat F_{\m\n} \\
& P^{\bot \, \kappa}_{\sigma}(q)
  P^{\bot \, \lambda}_{\rho}(-q) V_{\kappa}^{a \, (0)}(q) A_{\l}^{b\,(0)}(-q) \psi^{V}(q,z)' \psi^{A}(-q,z)
\end{split}
\ee
where prime denotes the derivative with respect to z. By differentiating with respect to the sources twice we find

\be
w_T(q^2)=-{2 N_c \over q^2} \int_{0}^{z_{\Lambda}} dz \, e^{- {1 \over
  2}\mu^2 \tau(z)^2} \psi^{V}(q,z)' \psi^{A}(-q,z)
\ee
where $\psi^{V}(q,z)$, $\psi^{A}(-q,z)$ are the solutions of bulk equations
of motion of the vector and axial vector  gauge fields, Eqs.(4.6) and
(4.9) of \cite{ikp2}.
We calculate the correlator by using the numerical solutions of
the vector and axial vector equations of motion.
In region  of low momenta, the result for $w_T(q^2)$ approximately matches  the relation
\be
w_T(q)={N_{c} \over q^2}-{N_c \over f_{\pi}^2}
\left[\Pi_A(q)-\Pi_V(q)\right]
\label{wtrellowq}
\ee
which was proposed in \cite{Son:2010vc},  as it is explained below.

\begin{figure}[ht]
\centering	
\includegraphics[width=.48\textwidth]{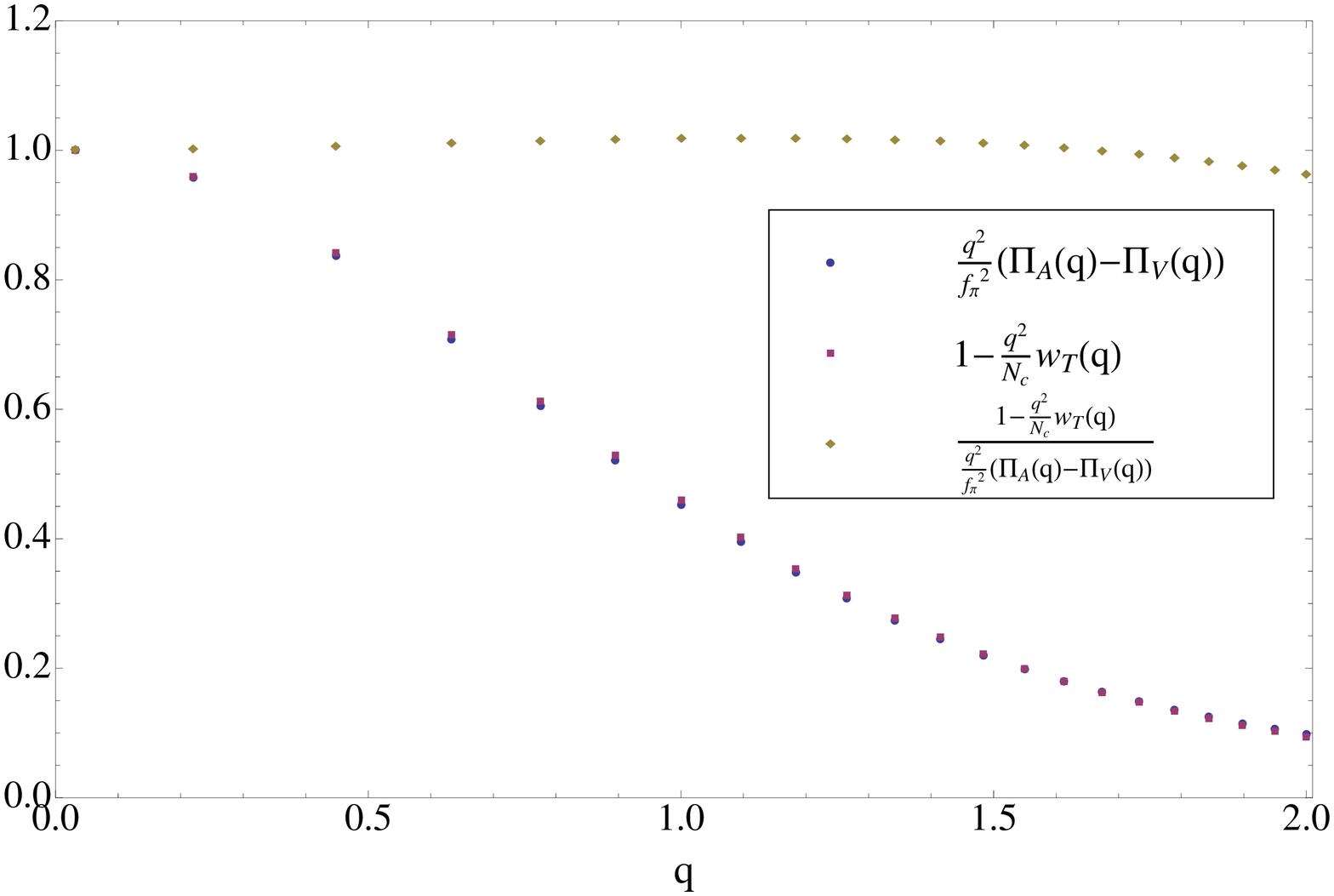}
       \includegraphics[width=.48\textwidth]{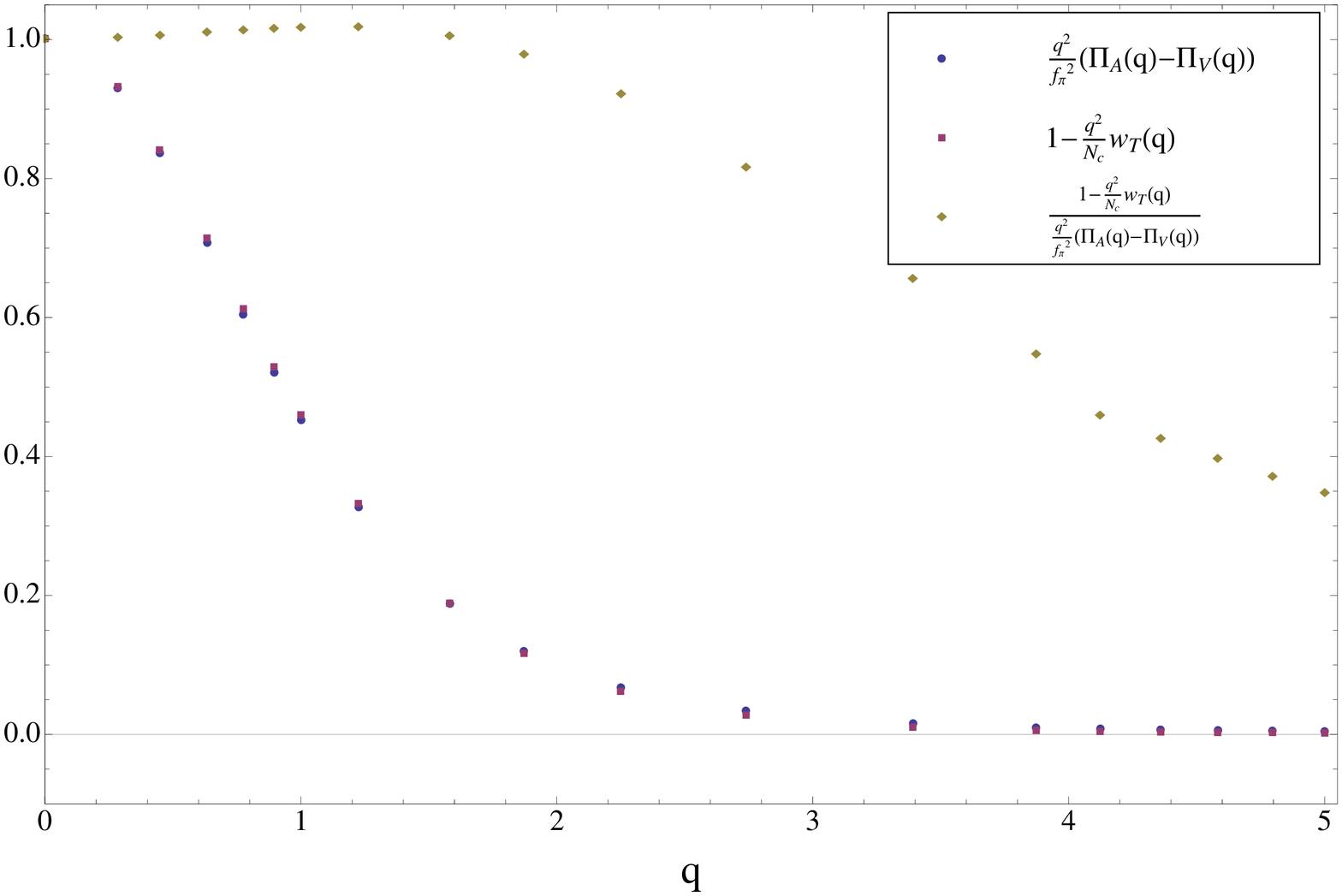}
	\caption{ The numerical result of $1-{q^2 \over N_c} w_T$, the difference of
          the vector and axial vector two-point functions $\left({q^2 \over f_{\pi}^2}
\left[\Pi_A(q)-\Pi_V(q)\right]\right)$ that appear
          in  Eq.({\protect\ref{wtrellowq}}), and their ratio are plotted a a function of $q$, which
          is measured in units of $z_\Lambda^{-1}$ which is essentially $\Lambda_{QCD}$, defined in \cite{ikp1}.
          On the left, there is a
          zoom of the plot for small $q$ and on the right the whole
          plot is depicted until large values of $q$.
	}
	\label{wsmallqcorr}
\end{figure}

On the left side of Fig.(\ref{wsmallqcorr}), the quantities $1-{q^2 \over N_c} w_T$, ${q^2 \over f_{\pi}^2}
\left[\Pi_A(q)-\Pi_V(q)\right]$ and their ratio are plotted in terms of
$q$ for low momentum. We observe that the ratio of the two quantities
is very close to one, so they coincide for
small values of momentum, namely $q\lesssim 2 \Lambda_{QCD}$.
 Hence Eq.(\ref{wtrellowq}) is satisfied in the low
momentum limit. The dependence of $w_T$ on momentum in this limit
matches the chiral perturbation theory analysis,
\cite{Knecht:2003xy},

\be
1-{q^2 \over N_c} w_T^{(QCD)}(q^2) \sim 1 -{q^2 \over N_c} 128 \pi^2
C_{22}^W+ {\mathcal O}(q^4) \;\;\;,\;\; q^2 \rightarrow 0
\label{cw}\ee
where $C_{22}^W$ is a coupling constant of the parity odd sector of the low energy chiral Lagrangian, \cite{Bijnens:2001bb}.
However, there is no independent calculation of the value of $C_{22}^W$
in order to verify Eq.(\ref{wtrellowq}) from the QCD viewpoint,
\cite{Knecht:2011wh}.  On the right side of figure \ref{wsmallqcorr}, we observe that as $q\gtrsim 2 \Lambda_{QCD}$ ,  the two quantities start differing substantially.

\begin{figure}[ht]
\centering	
\includegraphics[width=.55\textwidth]{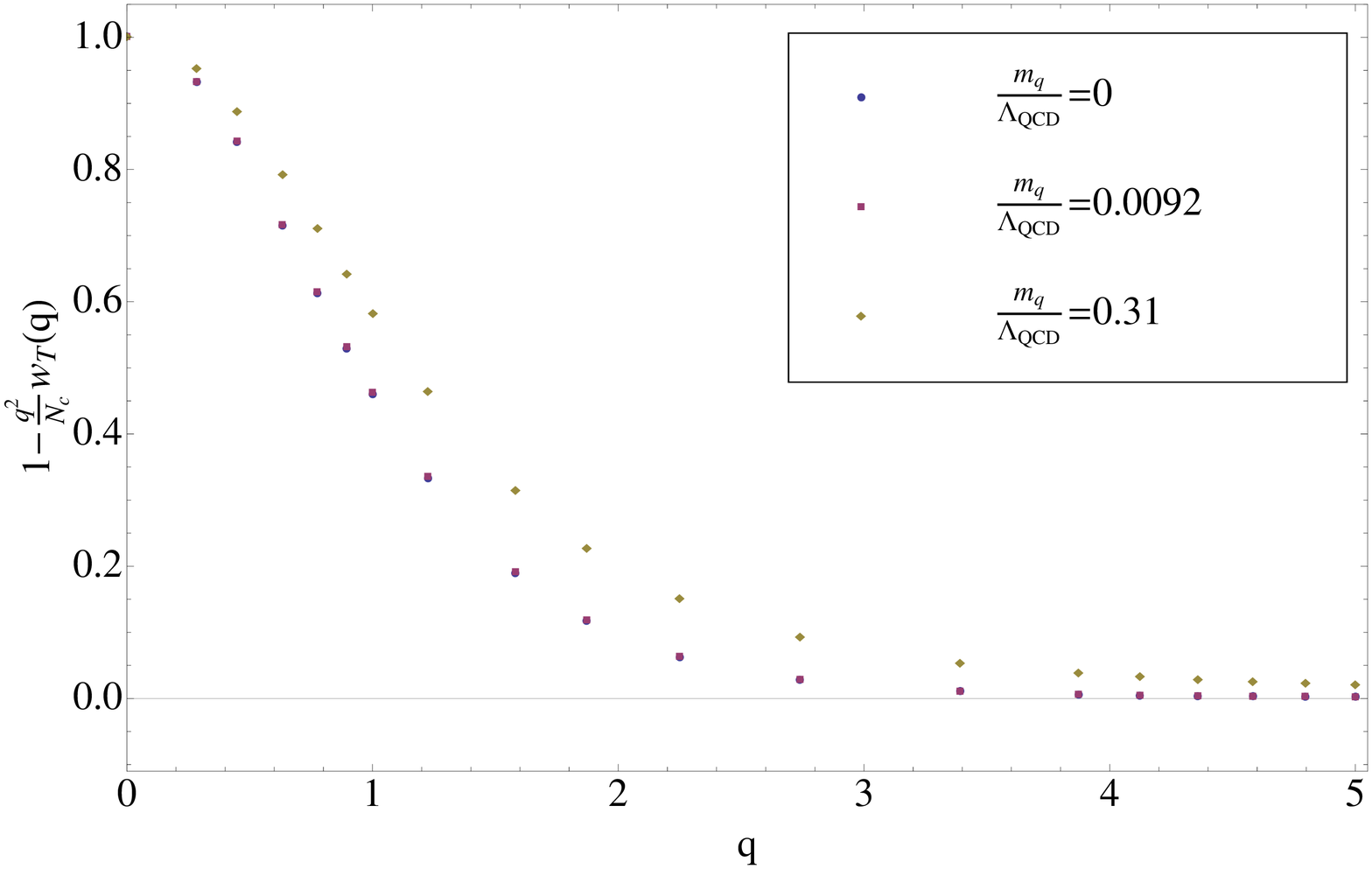}
	\caption{ The transverse part of the vector-axial vector
          correlator is plotted in terms of q for different bare quark
          masses. We used the values for the quark mass which
          were found by matching to the meson spectrum in
          \cite{ikp1} and \cite{ikp2}.
	}
	\label{wsmallqcorr2}
\end{figure}

In Fig.(\ref{wsmallqcorr2}), we plot the transverse part of the
correlator, $1-{q^2 \over N_c} w_T(q)$ for different values of the bare
quark mass. The slopes of the curves in Fig.(\ref{wsmallqcorr2}) for
low momenta lead to  the value of $C_{22}^W$ for different quark masses

\begin{equation}
\begin{split}
&C_{22}^W=6.71 \; 10^{-3} GeV^{-2} \;\;\; \mbox{for} \;\;\; {m_q\over \Lambda_{QCD}}=0 \\
&C_{22}^W=6.21 \; 10^{-3} GeV^{-2} \;\;\; \mbox{for} \;\;\; {m_q\over
  \Lambda_{QCD}}=0.0092\\
&C_{22}^W=4.45 \; 10^{-3} GeV^{-2} \;\;\; \mbox{for} \;\;\; {m_q\over \Lambda_{QCD}}=0.31
\end{split}
\end{equation}
where $C_{22}^{W}$ was defined in (\ref{cw}) and captures the low momentum asymptotics of the correlator. We have used $\Lambda_{QCD}=549 \; MeV$, as found by the fit
to meson spectra in \cite{ikp2}.
The mass ${m_q\over
  \Lambda_{QCD}}=0.0092$ corresponds to ${m_u+m_d\over 2}$ as fit in \cite{ikp2}.
The mass $ {m_q\over \Lambda_{QCD}}=0.31$ corresponds to the mass of the strange quark again from the same fit.

\begin{figure}[ht]
\centering
	\includegraphics[width=.55\textwidth]{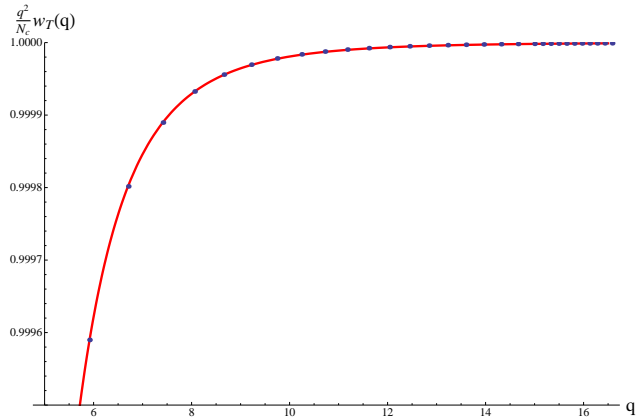}
	\caption{
	The numerical plot and the fit of the transverse part of the vector-axial
        vector correlator times $q^2/N_c$ in terms  $q$, for large Euclidean momentum. The function
        that fits the numerical data is $1-{18.75 \over
          q^{5.9974}}$.   As usual $q$ is given in units of $\Lambda_{QCD}$ as defined in \cite{ikp1}.
	}
	\label{wtfigfull}
\end{figure}

The transverse part of the vector-axial vector correlator times $q^2/N_c$, for large momentum, is plotted in Fig.(\ref{wtfigfull}). We have not been able to derive analytically the large $q$ asymptotics of this correlator.
We have therefore calculated it numerically and fitted a power law at sufficiently large $q$.
As seen in  Fig.(\ref{wtfigfull}), the subleading UV behavior of ${q^2
  \over N_c} w_T(q)$ is $1/q^6$, instead of the $1/q^4$ expected from QCD. This is not unexpected, and suggests that in the UV of the model we are using there are lighter stringy states (before the antisymmetric tensor) that contribute to the correlator and dominate its UV asymptotics.

We observe that the subleading
behavior of the correlator is subleading to the expected QCD
result. In \cite{Knecht:2002hr}, \cite{Czarnecki:2002nt} the
nonperturbative effect to the correlator was found by using operator
product expansion as it is mentioned in the introduction. It was shown that for large momentum, the
subleading part of $w^{QCD}_T$ is expected to be $\sim 1/q^6$, hence ${q^2 \over N_c}
w^{QCD}_{T} \sim 1/ q^4$. By finding the best fit to
our numerical data we find that the subleading term of ${q^2 \over N_c} w_T$ is
$\sim 1/q^{5.9974}$.
This disagreement with QCD for large
$q$ is suggesting that for the simplistic glue theory we are using , the fermionic operator in question does not appear in the appropriate OPE of the currents.

As it is shown in the next section the difference of the vector two point
function from the axial one for large momentum is found to be similar
to the QCD result which is $~ 1/q^6$, hence ${q^2 \over
  N_c}(\Pi_A(q^2)-\Pi_V(q^2) \sim 1/q^4$. Therfore, Eq.(\ref{wtrellowq}) is
violated for large momentum since our result for the vector-axial vector correlator
gives $1-{q^2 \over N_c} w_T \sim 1/q^{5.9974}$. A discussion of  Eq.(\ref{wtrellowq}) in the
context of QCD exists in \cite{Knecht:2011wh}.  The above outcome is
similar to the result of the hard wall AdS/QCD model which is analyzed in
Appendix B of \cite{Son:2010vc}.

\section{Vector and Axial current-current correlators}

We present here the vector and axial current-current correlators for
large Euclidean momentum, following \cite{Erlich:2005qh}, \cite{Da
  Rold:2005zs}. Those have already been calculated analytically in the abelian
case of our model in \cite{ikp2} and found to contain a subleading power of $1/q^4$, (for more details see  Eqs.(4.18) and (F.11) of \cite{ikp2}) that is absent from QCD which predicts a $1/q^6$ subleading behavior. However it turns that the analytical approximations made were unreliable. Here we will calculate the difference numerically and show agreement with QCD.

The two-point functions are defined as

\be
\int d^4 x d^4 y e^{i q x+i p y} \langle T \lbrace
J^{a}_{(V/A) \, \m} (x) J^{b}_{(V/A) \, \n} (y) \rbrace
\rangle=\delta^{ab}(q^2 \eta_{\m\n}-q_{\m} q_{\n}) \Pi_{V/A}(q^2) (2
\pi)^4 \delta^{(4)}(p+q)
\ee

\begin{figure}[ht]
\centering
	\includegraphics[width=.55\textwidth]{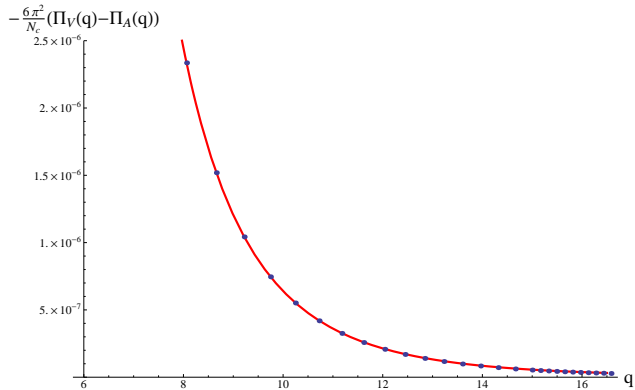}
	\caption{
	The numerical plot and the fit of the difference of the vector
        and axial vector 2 point functions. The fit reads ${0.653 \over q^{6.00933} }$.
        As usual $q$ is given in units of $\Lambda_{QCD}$ as defined in \cite{ikp1}.
	}
	\label{pivpia}
\end{figure}

We now calculate $\Pi_A(q^2)-\Pi_V(q^2)$ numerically using
the full equations of motion without any approximation. The result for
low momentum is shown in Fig.\ref{wsmallqcorr}. The
above difference is plotted in terms of $q$, Fig.(\ref{pivpia}), for
large Euclidean momentum and we find that the function that
fits the data is ${0.653 \over q^{6.00933} }$. The difference of the
axial vector and vector two point functions was calculated in QCD,
using operator product expansion method, \cite{Shifman:1978by}.   In the chiral limit the
QCD result reads

\be
\Pi_A(q^2)-\Pi_V(q^2)\sim{2 \pi \alpha_s\over q^6} \langle(\bar \psi_L
\gamma_{\mu} t^a \psi_L)(\bar \psi_R
\gamma_{\mu} t^a \psi_R)) \rangle
\ee
whose leading power agrees with the numerical result of our calculation at large momenta.

\section{Conclusions}

In the present work, we have shown that the relation (\ref{wtrellowq}) which was
proposed by Son and Yamamoto relating the vector-axial vector flavor current correlator in
weak electric field to the difference of the vector and axial vector
two point functions is valid only for low Euclidean momenta in the context of
an AdS/QCD model with non-trivial dynamics for the chiral condensate.

For large momenta,  this relation is no longer valid in our model. Moreover, the dependence of the
$w_T$ correlator on $q^2$ for large $q^2$ is subleading to that expected from QCD.

  We also notice that the difference of
the vector and axial vector flavor current two-point functions in our model falls off
as $q^{-6}$ for large $q^2$ as it is expected from  the operator
product expansion in QCD.

This computation could be sharpened by using a more realistic theory for the glue sector, like the Improved holographic QCD model, \cite{ihqcd}. This investigation in underway.

\section{Acknowledgements}\label{ACKNOWL}

We would like to thank A. Paredes for discussions.

This work was  partially supported by European Union grants
FP7-REGPOT-2008-1-CreteHEPCosmo-228644 and
PERG07-GA-2010-268246. I. Iatrakis work was supported by the project "HERAKLEITOS II - University of Crete" of the Operational Programme for Education and Lifelong Learning 2007 - 2013 (E.P.E.D.V.M.) of the NSRF (2007 - 2013), which is co-funded by the European Union (European Social Fund) and National Resources.

\section*{Note added}\label{ACKNOWL}

After this paper was in its last stages, \cite{new} appeared that studied the same correlator in the soft wall model.

\end{document}